%% file: main.tex
\documentclass[lettersize,journal]{IEEEtran}

\usepackage{booktabs} 
\usepackage{orcidlink}
\usepackage{url} 
\usepackage{tabularx}
\usepackage{cite}
\usepackage[numbers, sort]{natbib}
\usepackage{multirow}
\usepackage{tcolorbox}
\usepackage{amsmath,amssymb,amsfonts}
\usepackage{algorithmic}
\usepackage{graphicx}
\usepackage{textcomp}
\usepackage{xcolor}
\def\BibTeX{{\rm B\kern-.05em{\sc i\kern-.025em b}\kern-.08em
    T\kern-.1667em\lower.7ex\hbox{E}\kern-.125emX}}

\usepackage [autostyle, english = american]{csquotes}
\MakeOuterQuote{"}
\usepackage{hyperref}
\hypersetup{
    colorlinks,
    linkcolor={red!50!black},
    citecolor={blue!50!black},
    urlcolor={blue!80!black}
}
\usepackage[dvipsnames]{xcolor}

\newcommand{\ourFramework}{FuzzSight}

\newcommand{\inlineheading}[1]{\vspace{2mm}\noindent\textbf{#1}}

\newcommand{\textbox}[1]{
    \begin{tcolorbox}[colback=gray!5!white,colframe=gray!100!black]
        #1
    \end{tcolorbox}
}
\newcommand{\needAttention}[1]{\textcolor{Black}{#1}}

\begin{document}

\title{Enhancing Code Review through Fuzzing and Likely Invariants}

\author{Wachiraphan Charoenwet~\orcidlink{0000-0002-9814-3514},
        Patanamon Thongtanunam~\orcidlink{0000-0001-6328-8839},
        Van-Thuan Pham~\orcidlink{0000-0002-9871-3695},
        and Christoph Treude~\orcidlink{0000-0002-6919-2149}
    \thanks{W.~Charoenwet, P.~Thongtanunam, and V.T.~Pham are with The University of Melbourne, Victoria, Australia, 3000\\
    E-mail: \{charoenwetw, patanamon.t, thuan.pham\}@unimelb.edu.au}
    \thanks{C.~Treude is with Singapore Management University, Singapore, 188065\\
    E-mail: ctreude@smu.edu.sg}
    
}

\maketitle

\begin{abstract}


\needAttention{Many software projects employ manual code review to gatekeep defects and vulnerabilities in newly introduced code before integration. 
However, reviewers often work under time pressure and rely primarily on static inspection of code changes, leaving the dynamic aspects of program behavior largely unexplored during review.
Dynamic analyses could help reveal such behaviors, but they are rarely integrated into review practices.
Among them, fuzzing is typically applied later in the testing pipeline to uncover crashing bugs.
Yet its ability to exercise code with diverse inputs also makes it promising for exposing non-crashing, but unexpected, behaviors earlier in the review process.
Still, without suitable mechanisms to analyze program behaviors, the rich data produced during fuzzing remains inaccessible to reviewers, limiting its practical value in this context.
}
  

\needAttention{We hypothesize that unexpected variations in program behaviors could signify potential bugs.
To capture such variations, we look beyond static inspection and leverage dynamic execution data, which fuzzing readily produces in abundance.
Instead of requiring reviewers to infer potential behavioral consequences solely from source code and static analyses, the impact of code changes can be automatically captured at runtime.
Representing program behavior as likely invariants, dynamic properties consistently observed at specific program points, can provide practical signals of behavioral changes, potentially revealing unintended or anomalous behavior.
Such signals offer a way to highlight code regions exhibiting behavioral differences and to distinguish between intended changes and unexpected behavioral shifts.}

We present \ourFramework{}, a framework that leverages likely invariants from non-crashing fuzzing inputs to highlight behavioral differences across program versions.
By surfacing such differences, it provides insights into which code blocks may need closer attention.
In our evaluation, \ourFramework{} flagged 75\% of regression bugs and up to 80\% of vulnerabilities uncovered by 24-hour fuzzing.
It also outperformed static security application testing (SAST) in identifying buggy code blocks, achieving ten times higher detection rates with fewer false alarms.
In summary, \ourFramework{} demonstrates the potential and value of leveraging fuzzing and invariant analysis for early-stage code review, bridging static inspection with dynamic behavioral insights.

\end{abstract}

\section{Introduction}
\label{introduction}
\input{section/1-introduction-new}


\section{Background and Related Work}
\label{background}
\input{section/2-background}

\section{\ourFramework{} Design}
\label{design}
\input{section/3-design}

\section{Study Settings}
\label{study_settings}
\input{section/4-study}

\section{Analysis and Results}
\label{results}
\input{section/5-result}

\section{Discussion}
\label{discussion}
\input{section/6-discussion}


\section{Threats to Validity}
\label{threats_to_validity}
\input{section/8-threats}

\section{Conclusion}
\label{conclusion}
\input{section/9-conclusion}

\section{Acknowledgements}
\label{acknowledgements}
\input{section/10-acknowledgements}

\bibliography{references.bib}{}
\bibliographystyle{IEEEtran}

\end{document}

%% file: section/1-introduction-new.tex

Software systems should be inspected and tested early to minimize the impact of late bug discovery~\cite{Stecklein2004ErrorCycle, Karg2011AResearch}. 
Code review, a common early-stage quality assurance practice, aims to detect potential issues in proposed changes before integration.
However, effective code review requires deep code understanding, which is both time-consuming and cognitively demanding~\cite{Pascarella2018InformationReview}.
In practice, reviewers often face time constraints~\cite{Kononenko2016CodeIt}, leading to reviews that focus on surface-level aspects while overlooking more subtle problems.
\needAttention{
Automated program analysis tools, such as static analyzers, can reduce reviewer effort by generating warnings similar to reviewer comments~\cite{Singh2017EvaluatingEffort}, suggesting potential defects worth addressing~\cite{Panichella2015WouldReviews}, or prioritizing code changes under reviews~\cite{Charoenwet2024AnReview}.
Yet, static analysis tools primarily rely on predefined rules and syntactic patterns, which limit their ability to account for dynamic behaviors, complex control-flow interactions, and subtle context-dependent effects.}
As a result, they struggle to capture the behavioral properties of programs~\cite{Ayewah2008UsingBugs} that are critical for reasoning about the implications of code changes.

\needAttention{
In contrast, dynamic analysis, particularly fuzzing, offers a complementary perspective by executing the program under diverse, automatically generated inputs.
In traditional software development workflows, fuzzing is usually applied in the later stages of the SDLC, after code review and integration, where its findings, such as crashes or assertion violations, primarily guide security testing and post-integration quality assurance by security experts~\cite{Klooster2023ContinuousPipelines}.
However, this rich behavioral information could also be valuable during code review, if presented in a way that helps reviewers reason about the consequences of code changes.
For example, differences in a program's behaviors are often more relevant to understanding the implications of code changes, yet such insights are rarely extracted or highlighted during the review process.
Even when artifacts such as execution traces are collected, they tend to be too technical and low-level for practical use by reviewers.
Thus, the challenge remains in adapting fuzzing's strengths, its ability to reveal behavioral diversity, into concise information that can meaningfully assist review activities.
}

\needAttention{
One promising way to make this information available is to focus on behavioral deviations that indicate potential issues.
Many bugs, especially regression bugs, manifest as behavioral deviations rather than outright crashes or failures.
Detecting such deviations can help identify suspicious code changes responsible for unexpected effects.
Our intuition further suggests that security vulnerabilities may also surface through distinctive behavioral patterns observable at runtime.
Likely invariants, properties that consistently hold at specific program points for a given input~\cite{Ernst2007TheInvariants}, provide a structured way to capture and summarize these behaviors.
We hypothesize that differences in likely invariants across program versions can signal anomalous behaviors introduced by code changes~\cite{Menarini2017Semantics-assistedStudy}.
Fuzzing can produce inputs that reveal rich execution data, capturing program behavior in the form of likely invariants even without triggering crashes.
This motivates our approach, which focuses on analyzing these behavioral differences to provide behavioral insights for code review.
}


\needAttention{In this work, we develop \ourFramework{}, a framework that leverages fuzzing inputs to capture program behaviors as likely invariants and compares them across two program versions.
By focusing on non-crashing executions early in the review process, the framework can detect behavioral differences that are often invisible to traditional fuzzing or static analyzers applied later in the development pipeline.
As a result, \ourFramework{} highlights behavioral differences, pinpointing the suspicious code blocks based on the differences detected, distinguishing harmless refactorings from unintended regressions, and potentially prioritizing code changes during the code reviews.
}

To explore the feasibility of leveraging non-crashing inputs for likely invariant analysis, we evaluate \ourFramework{} through a multi-stage experiment, starting with a motivating example and a preliminary experiment to validate our hypothesis.
Next, we evaluate the effectiveness of \ourFramework{} in detecting two types of bugs: vulnerabilities and regressions, that can manifest as subtle behavioral shifts without triggering failures. 
Vulnerabilities may evade code reviews and fuzzing due to complexity or lack of oracles, while regressions often arise from mistakes during code modification. 
These traits align with \ourFramework{}’s use of likely invariants to detect behavioral deviations.
Finally, we evaluate the performance of \ourFramework{} against static application security testing (SAST) tools' warnings in detecting buggy code changes at the code block level.

Our empirical results show that likely invariant differences occur in 10\%-58\% of vulnerabilities, and notably, up to 80\% of bugs that fuzzing tools cannot expose in 24 hours.
For real-world regression bugs, \ourFramework{} flags 75\% of bugs reached by fuzzing inputs.
\ourFramework{} can also flag buggy code blocks in ten times as many bugs as SAST warnings, highlighting its potential for fine-grained bug detection in code reviews with a lower false alarm rate.
Based on the results, \ourFramework{} demonstrates the feasibility of integrating fuzzing and invariant analysis into early-stage code review, bridging the gap between static inspection and dynamic behavioral understanding.

\inlineheading{Novelty and Contribution:} This work offers the following novelty and contributions:

\begin{enumerate}
    \item \textit{Technique}: Leverage non-crashing fuzzing inputs to identify behavioral shifts via likely invariant analysis, enabling fuzzing to inform code review decisions.
    \item \textit{Empirical Findings}: \textcircled{\footnotesize1} Investigate benefits of leveraging fuzzing to detect vulnerability-related bugs and bug-inducing regression code changes. \textcircled{\footnotesize2} Evaluate the detection of likely invariant differences in buggy code changes against SAST warnings.
    \item \textit{Framework}: Develop likely invariant analysis framework that can integrate with state-of-the-art fuzzing tools.
\end{enumerate}

\noindent \ourFramework{} and experiment replication package are available at: \url{https://doi.org/10.5281/zenodo.15458082}


\inlineheading{Paper Organization:} The rest of this paper is organized as follows. 
Section~\ref{background} introduces background and related work. 
Section~\ref{approach_overview} presents a motivating example and preliminary experiments. 
Section~\ref{design} details the framework design. 
Section~\ref{study_settings} describes the experimental setup, and Section~\ref{results} reports the results. 
Section~\ref{discussion} discusses the implications of our findings. 
Section~\ref{threats_to_validity} outlines threats to validity, and Section~\ref{conclusion} concludes the paper.

%% file: section/2-background.tex


This section provides background on shift-left security assurance principle and the role of fuzzing in shift-left, and reviews related work on differential-testing approaches for detecting bugs from behavioral inconsistencies and on likely invariants as a means to capture program behaviors.

\subsection{Shift-left Quality Assurance}
The shift-left principle encourages identifying and addressing various issues early in the software development lifecycle~\cite{Firesmith2015FourTesting, Dawoud2024BetterDevelopment}. 
Code review aligns with this principle by allowing reviewers to inspect code changes made by developers before integration. 
Through this process, reviewers can identify potential issues and suggest improvements that enhance both functionality and maintainability of the code~\cite{Mantyla2009WhatReviews}. 
Despite its benefits, code review is often constrained by limited time and human effort. 
As a result, reviewers tend to rely heavily on static inspection, reading and reasoning about code without executing it~\cite{Spadini2018WhenTests, Ebert2021AnReviews, Kononenko2016CodeIt}. 
This static nature means that dynamic aspects of program behavior, such as runtime state changes or subtle control-flow dependencies, can be overlooked.
Consequently, bugs that involve dynamic behavioral issues can remain undetected, especially when they manifest only under specific runtime conditions. 

\begin{figure}[t]
  \centering
  \includegraphics[width=1\linewidth, scale=1]{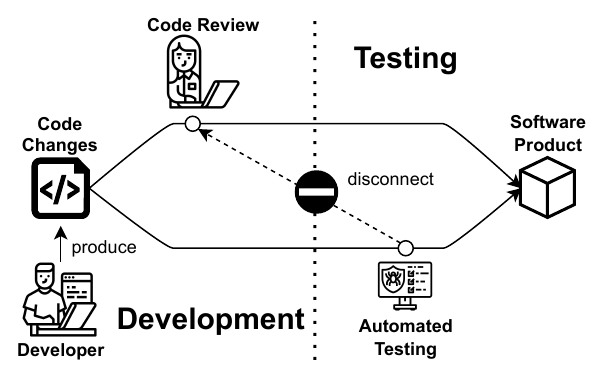} 
  \caption{In typical software development lifecycle, code review and automated testing can operate in parallel, yet remain disconnected.
  }
  \label{fig:background-sdlc}
\end{figure}

\needAttention{
On the other hand, various automated testing techniques, including unit tests, integration tests, and fuzzing are typically executed after code changes are committed, often in the later stages of the CI/CD pipeline~\cite{Deshmukh2025AutomatedPipeline, Klooster2023ContinuousPipelines}.
As shown in Fig.~\ref{fig:background-sdlc}, code review occurs in the development stage, while automated testing may run in parallel or afterward in subsequent stages, creating a temporal and process disconnect between the two activities.
Although automated tests are effective at exercising program behavior and checking correctness, the results are rarely presented in a way that can inform reviewers about the runtime impact of code changes.
As a result, the insights from automated testing are often not readily available to reviewers, highlighting an opportunity to better integrate runtime behavioral information into the review process~\cite{Panichella2020AnReview}.
}

\subsection{Fuzzing}
\needAttention{While integrating quality assurance early in the development lifecycle can reduce latent issues, dynamic analysis techniques such as fuzzing are still predominantly applied in later stages~\cite{Klooster2023ContinuousPipelines, Dawoud2024BetterDevelopment}.
Fuzzing is a form of automated dynamic testing that executes programs with systematically or randomly generated inputs while monitoring for unwanted behaviors using fuzzing oracles~\cite{Barr2015TheSurvey}.
It can uncover both functional bugs and security vulnerabilities in complex software systems.
By generating diverse inputs and observing program responses, fuzzing can reveal unintended behaviors and subtle execution paths that may compromise program integrity, highlighting scenarios where inputs could be exploited to influence the program in unexpected ways~\cite{Theisen2015ApproximatingTraces}.}

\needAttention{
Fuzzing has proven effective at detecting bugs and security vulnerabilities~\cite{Liang2018Fuzzing:Art}, but it often requires extended run time to explore complex execution paths~\cite{Liyanage2023ReachableFuzzing}.
In fast-paced development environments such as Continuous Integration and Deployment (CI/CD), this requirement can limit fuzzing’s ability to uncover bugs introduced by recent code changes.
Even fuzzers designed for CI/CD, such as AFLChurn~\cite{Zhu2021RegressionFuzzing}, which targets modified code, face challenges as they may prioritize coverage while missing subtle interactions or data-flow effects around the changes.
As a result, some bugs remain undetected even when the affected code is executed~\cite{Fioraldi2021TheFuzzers}, particularly those that manifest as behavioral differences rather than crashes observable by standard oracles~\cite{Pan2024EDEFuzz:Exposures}.
This highlights the need for complementary techniques to identify latent issues that fuzzing alone may miss.
}


\needAttention{
To support early-stage quality assurance and code review, it is imperative to maximize the utility of fuzzing by leveraging its by-products, such as non-crashing inputs, to uncover bugs that might otherwise go undetected.
These fuzzing-generated inputs can reveal new program behaviors through techniques like automated debugging~\cite{Parnin2011AreProgrammers} and support more thorough testing of edge cases that may be overlooked~\cite{Pascarella2018InformationReview, Panichella2020AnReview}. 
The abundance of inputs provides a richer view of program behavior, which can help identify unintended effects introduced by code changes. 
This potential makes fuzzing-generated inputs a promising resource for shift-left quality assurance and for supporting human code review.
}

\subsection{Differential Testing-Based Approaches}
\needAttention{
\ourFramework{} aims to pinpoint suspicious code blocks by detecting unexpected shifts in program behaviors. 
While the intuition is related to Differential Testing (DT), our approach diverges in two key aspects.
First, instead of relying on manually crafted or purposefully designed test cases, \ourFramework{} leverages fuzzing to generate diverse execution traces that more realistically capture diverse program behaviors.
Second, \ourFramework{} operates at the code-block level, enabling automated and fine-grained detection of behavioral shifts directly associated with specific code changes.
By contrast, DT techniques typically compare program outputs across program implementations, using behavioral discrepancies as a cross-reference oracle for bug detection~\cite{Gulzar2019PerceptionTesting}.
}

\needAttention{
Although DT techniques are effective for uncovering unknown bugs in the production stage, they are less suitable for code review because they typically report behavioral differences without identifying where in the code they originate.
This limits their usefulness for reviewers, who need guidance on the problematic code lines or code blocks.
Prior work proposed GETTY~\cite{Menarini2017Semantics-assistedStudy}, a DT-based approach for localizing buggy functions, which provides finer detail than traditional DT but still operates at the coarse level of functions.
GETTY also requires reviewers to manually examine the results to interpret how functions’ behaviors have changed and how those changes could map back to code, potentially adding cognitive load and contributing to review fatigue~\cite{Pascarella2018InformationReview}.
Moreover, the most recent DT techniques are domain-specific and do not address the fine-grained, domain-agnostic needs of general code review.
For instance, TWINFUZZ~\cite{Leonelli2025TWINFUZZ:Stacks} focuses on detecting discrepancies between software and hardware video decoders, DiffCSP~\cite{SeongilWi2023DiffCSP:Testing.} compares how browsers process generated Content Security Policy and HTML instances, and Li and Rigger~\cite{Li2024FindingTesting} test differential behaviors across XML processors.
}

\needAttention{
\subsection{Likely Invariants and Their Applications}
Program invariants provide a simplified way to capture and represent program behaviors~\cite{Barr2015TheSurvey, Menarini2017Semantics-assistedStudy}.
They represent properties that are always true at certain program states.
For example, 
\texttt{x > 0} or
\texttt{input\_a} is not null.
Program invariants can be used for debugging, testing, verification, and bug detection by ensuring expected behavior during execution~\cite{Barr2015TheSurvey, Fioraldi2021TheFuzzers, Le2016AInvariants}.
}

Typically, program invariants are derived from formal specifications.
When specifications are absent, \textit{likely invariants} can be inferred through concrete executions of the program using valid inputs.
They are approximations of program invariants, offering an alternative representation of program behavior~\cite{Ernst2007TheInvariants}.
Likely invariant generators can infer properties, such as variable relationships and value constraints similar to program invariants, based on values observed during execution.
Therefore, likely invariants offer a practical substitute for program invariants when formal specifications are unavailable.


Previous research has explored applications of likely invariants.
Le et al.~\cite{Le2016AInvariants} proposed Savant, an approach that leverages likely invariants to localize failure causes at the method level.
While Savant analyzes differences in likely invariants, it assumes failures are already known via a test oracle, a condition that may not hold in code review.
Fioraldi et al.~\cite{Fioraldi2021TheFuzzers} developed INVSCOV, a fuzzing tool that navigates fuzzing direction by prioritizing paths that exhibit changing likely invariants, showing promise in discovering new bugs in frequently fuzzed code.
However, its focus on steering fuzzing through control-flow-based invariant changes overlooks the complementary opportunity of systematically analyzing invariants over an existing seed corpus.

\needAttention{
These works highlight the potential of likely invariants to capture behavioral differences. 
Our work builds on this foundation by leveraging fuzzing-derived non-crashing inputs and invariant analysis to identify behavioral shifts between program versions at the block level, without relying on failing executions.
}

\section{Approach Overview}
\label{approach_overview}
This section provides an overview of FuzzSight, presenting a motivating example and a preliminary experiment that demonstrate its ability to analyze behavioral differences across program versions.

\subsection{Motivating Example}
\label{motivating_example_section}
\begin{figure}[t]
  \centering
  \includegraphics[width=1\linewidth, scale=1]{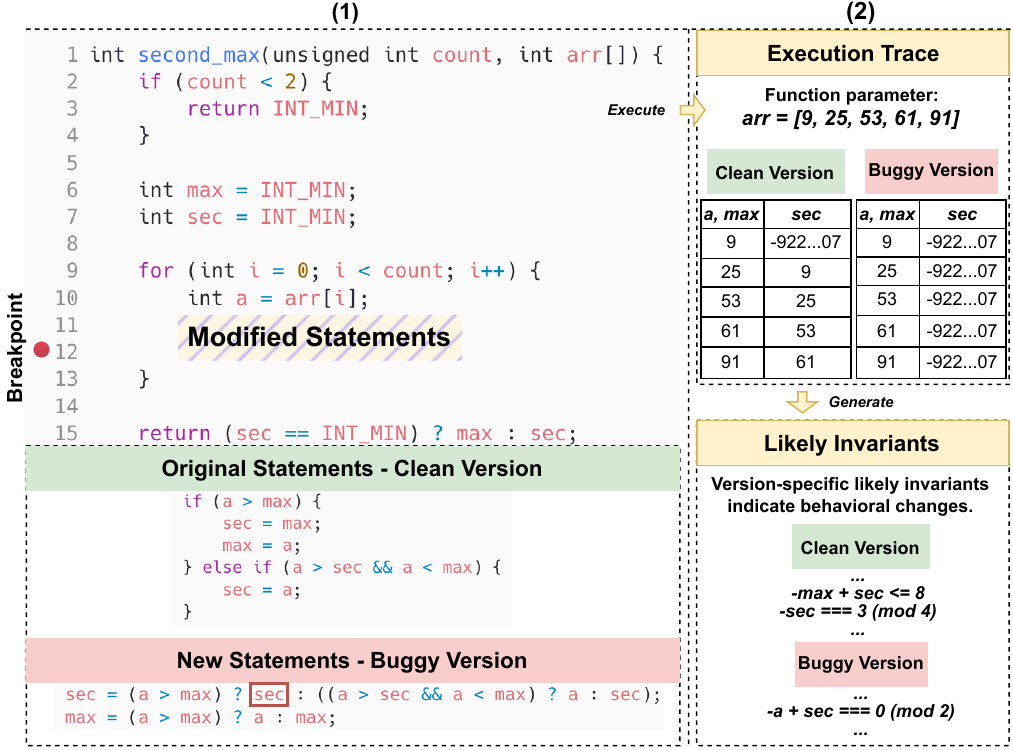} 
  \caption{Motivating example--\textcircled{\scriptsize 1} a bug is introduced when modifying a function (red box). \textcircled{\scriptsize 2} identical inputs yield different traces and likely invariants across versions. 
  }
  \label{fig:motivating-example}
\end{figure}

To motivate the potential of likely invariant analysis in changed code, we present a small program that introduces a regression bug (Fig.~\ref{fig:motivating-example}.\textcircled{\footnotesize1}).
The \textit{second\_max} program searches and returns the second-largest item in an array.
In the clean version, the \texttt{if-else} conditional statement was used. 
A developer then replaced a code block with a ternary operator for conciseness.
The automated static program analyzers did not detect any issues. 
However, when the program is executed, the function could fail to return the second-largest element, as the variable \texttt{sec} is incorrectly assigned to itself.

Given this example, it would be challenging for reviewers to carefully inspect and test the code changes during the development cycle.
Even for seemingly simple modifications, revealing the new behavior requires substantial effort.
It is a logical bug, or \textit{silent bug}, that does not alter the control flow or cause a crash.
Moreover, as it only produces incorrect results under specific test cases, it may go unnoticed due to \textit{accidental correctness}.
In this work, we argue that likely invariants can serve as adaptable indicators of these traces, enabling automated techniques to detect discrepancies.

In other words, this bug could be detected through execution and program behavior via dynamic analysis.
With the likely invariant analysis, we can detect the behavioral shifts, which could potentially reduce the reviewer's effort.
As illustrated in Fig.~\ref{fig:motivating-example}.\textcircled{\footnotesize2}, a single input produces different execution traces.
Analyzing execution traces directly is challenging due to their large volume and complexity, which hinders effective examination~\cite{Sanchez2019ASoftware}.
Instead, these execution traces can be used to infer the distinct sets of likely invariants that represent the program behaviors.
Given that both versions are expected to behave identically, this divergence can alert reviewers about potential unintended behavioral changes.

\subsection{Preliminary Experiment}

To further motivate this work, we evaluate the approach on small, well-known C programs from The Art of Computer Programming~\cite{Knuth2005TheProgramming}, including array concatenation, bubble sort, factorial, the greatest common divisor (GCD), permutation, second maximum search, and string reversal.
Each program is transformed into three versions: two clean, functionally equivalent versions and one buggy version.
The first version follows the original algorithm, while the second applies logic-preserving changes such as refactoring or an alternative algorithm.
The buggy version is derived from the second by deliberately altering its logic to introduce a behavioral bug.\footnote{Example programs are included in the replication package.}

We perform fuzzing and generate likely invariants, comparing likely invariants from the same inputs in two version pairs: \textcircled{\footnotesize1} the two clean versions and \textcircled{\footnotesize2} one clean and one buggy version.
We measure the difference using the Dice–Sørensen coefficient to represent the distance ($ 1-Dice $) between likely invariant sets from two versions.
A greater distance indicates more difference.
We determine the distribution of distances across inputs and identify the largest x-axis value where a local peak occurs, i.e., many inputs yield that distance. 
A peak at zero indicates no behavioral difference between versions.

\input{table/table-preliminary-experiment}

Table~\ref{table:preliminary-experiment} suggests that likely invariant differences can indicate behavioral changes. 
Clean versions consistently produce identical invariants, while the buggy version shows notable deviations across all examples.
Even in GCD and permutation, buggy-clean differences exceed those between clean versions, which result from alternative implementations. 
For GCD, the clean versions' differences stem from parameter swapping in recursion. 
For permutation, both clean versions use recursion, which can lead to execution trace differences~\cite{Ciobaca2023Operationally-basedLCTRSs}.


These differences in likely invariants between program versions highlight the potential of fuzzing inputs to support code reviews.
This approach 
pinpoints the area of code changes that contains behavioral changes, which reviewers should further investigate.
Unlike fault localization techniques~\cite{Le2016AInvariants}, where differences in likely invariants help expose known bugs, behavioral changes introduced by code changes do not necessarily indicate bugs.
For instance, behavioral changes can be intentional or expected as a result of code changes, e.g., permutation and the greatest common divisor in our preliminary experiment (Table~\ref{table:preliminary-experiment}).
Our approach focuses on identifying behavioral changes as potential indicators of unintended bugs.

%% file: table/table-preliminary-experiment.tex
\begin{table}[h]
\centering
\scriptsize
\caption{Preliminary Experiment--likely invariant differences between versions of example programs, measured by the largest distance with local peak in distributions across inputs.}

\begin{tabularx}{\linewidth}{
    >{\hsize=1.05\hsize}X |
    >{\centering\arraybackslash\hsize=1\hsize}X
    >{\centering\arraybackslash\hsize=0.95\hsize}X 
}
\toprule
\textbf{Example Program} &\textbf{ Clean 1 - Clean 2 } & \textbf{Clean 1 - Buggy } \\
 \midrule
\textbf{array concatenation} & No Difference & \textbf{0.81}  \\
\textbf{bubble sort} & No Difference & \textbf{0.58}  \\
\textbf{factorial} & No Difference & \textbf{0.99}  \\
\textbf{greatest common divisor} & 0.71\textsuperscript{\textdagger} & \textbf{0.91}  \\
\textbf{permutation} & 0.21\textsuperscript{\textdagger} & \textbf{0.41}  \\
\textbf{second maximum search} & No Difference & \textbf{0.02}  \\
\textbf{string reversal} & No Difference & \textbf{0.62}  \\

 \bottomrule
\multicolumn{3}{l}{\textdagger~Invariant differences are the result of alternative implementations. }
 
\end{tabularx}

\label{table:preliminary-experiment}
\end{table}

%% file: section/3-design.tex
\begin{figure*}[t!]
  \includegraphics[width=\linewidth, scale=1]{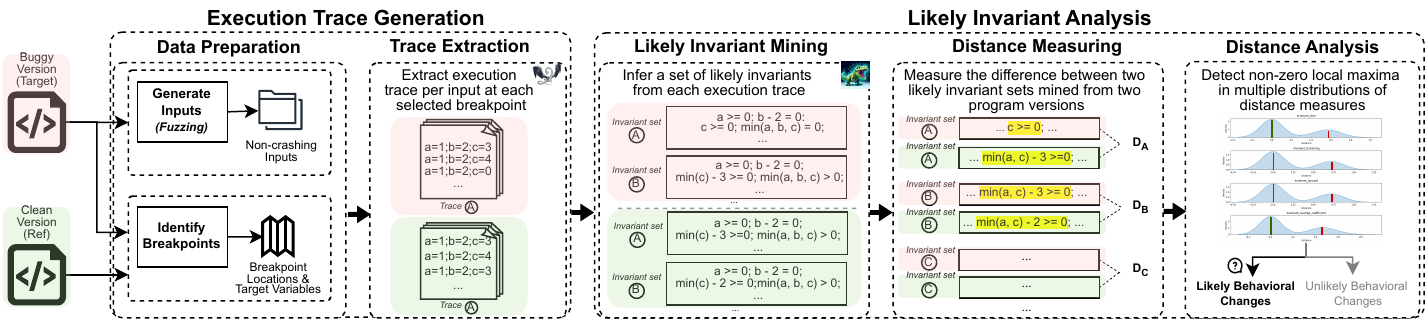} 
  \caption{\ourFramework{} Design. Execution traces from two program versions are collected using non-crashing inputs (e.g., from fuzzing tools). Likely invariants are then mined from each trace, and differences between the resulting likely invariant sets are measured to reveal behavioral changes. }
  \label{fig:overall-approach-design}
\end{figure*}


\ourFramework{} is a framework for analyzing data from dynamic analysis tools, aiming to help identify unexpected behavioral changes between two program versions: \textit{clean} and \textit{buggy}. 
The buggy version contains the \textit{target bug}, while the clean version is free from this bug.
The framework comprises two components (Fig.~\ref{fig:overall-approach-design}.): 
\textcircled{\footnotesize1} execution trace generation, and
\textcircled{\footnotesize2} likely invariant analysis.
We detail each component below.



\subsection{Execution Trace Generation}
Execution trace generation involves running the target program with an input and extracting values of variables at selected \textit{breakpoints}, i.e., specific locations in the target program where likely invariant differences are analyzed.
This comprises two steps: fuzzing and extracting execution trace.

\subsubsection{Fuzzing and Breakpoint Selection}
We use fuzzing to generate an input corpus.
In particular, we use inputs that do not crash the target program, i.e., \textit{non-crashing inputs}, from the fuzzing input queue.
These inputs can help expose the dynamic behaviors of the program, although they do not trigger crashes.
The code accessed by these inputs is considered high-risk due to the attack surface it presents~\cite{Theisen2015ApproximatingTraces}.

Breakpoints in the target program must be selected to generate the execution trace. 
In general, placing a breakpoint in a code block after the modified code is suitable for our framework, as it can capture the effects of the changes.
In practice, techniques such as program slicing can assist in identifying code lines impacted by changes.
Breakpoint selection strategies in our experiments are described in Section~\ref{experiment_dataset_preparation}.

\subsubsection{Trace Extraction}
We extract the values of variables at selected breakpoints from each input.
If the selected breakpoints are executed multiple times, we retain the values from all execution rounds.
This approach allows us to track changes in variable values.
We set breakpoints in the target functions and retrieve the values of variables until the execution is completed using an automated debugging script in \texttt{lldb}, the LLVM debugger~\cite{LLDB}.
To streamline trace extraction, we format all variable values as decimals.
String variables are converted into a decimal value using a deterministic hash algorithm to enable likely invariant mining.

\subsection{Likely Invariant Analysis}
This section describes likely invariant mining, distance measurement, and distance analysis across program versions.

\subsubsection{Likely Invariant Mining}
We employ likely invariants~\cite{Nguyen2022UsingInvariants}, i.e., properties that always hold at certain program locations, for analyzing how code changes introduce new behaviors.
We mine likely invariants with DIG~\cite{Nguyen2014DIG}, a lightweight tool for likely invariant inference, capable of generating likely invariants for various numerical relations such as linear and non-linear equalities as well as their corresponding inequalities, from raw source code and pre-formatted execution traces.
Compared to Daikon~\cite{Ernst2007TheInvariants}, a well-known invariant mining tool that typically requires source code and produces invariants only at function entry and exit points, DIG offers greater integration flexibility.
To enable likely invariant generation with DIG, we convert the execution traces into DIG’s trace format before initiating the mining process.
The resulting sets of invariants for an input, corresponding to both program versions, are then stored for subsequent analysis.

We suppress DIG's likely invariants simplification and filtering mechanisms because they can introduce randomness in the results~\cite{Nguyen2022UsingInvariants}.
Additionally, to manage the computational overhead of likely invariant mining, we impose a five-minute timeout per input for likely invariant mining.

\subsubsection{Distance Measurement}
We use set distances as a proxy for differences between likely invariant sets at a breakpoint in the two program versions under the same input.
Four distance measures are used: Dice-Sørensen coefficient, Hamming distance, Jaccard index, and Overlap coefficient.
Prior works used these distance measures to analyze various artifacts in software engineering research~\cite{Shtern2012ClusteringEngineering, Allyson2019SherlockCode, Taheri2020Similarity-basedFeatures}.
For example, Dice-Sørensen coefficient and Jaccard index are used in software clustering, Hamming distance is used for malware detection, and Overlap coefficient is used for measuring code similarity.
Except for Hamming distance, the other measures yield similarity scores in $[0,1]$; we subtract these from $1$ to interpret them as distances.
Additionally, when likely invariants are generated in only one version of the program, we treat them as the maximum distance.
Given sets $U$ and $V$ of likely invariants from each program version, distances are described as follows.

\begin{enumerate}
    \item \underline{Dice-Sørensen coefficient}: distance between two sets based on their common elements; $ 1 - \frac{2|U \cap V|}{|U| + |V|} $
    \item \underline{Hamming distance}: the number of differing positions between two equal-length sequences of elements; $ \sum_{i=1}^{k} |U_{i} - V_{i}| $
    \item \underline{Jaccard index}: distance of two sets by dividing the intersection by the union; $1 - \frac{|U \cap V|}{|U \cup V|} $
    \item \underline{Overlap coefficient}: distance of two sets based on their intersection relative to the smaller set;  $ 1 - \frac{|U \cap V|}{\min(|U|, |V|)} $
\end{enumerate}

We use multiple distance measures because we aim to capture and report various types of differences in likely invariant sets.
Hamming distance grows rapidly with positional differences. 
Jaccard and Dice-Sørensen reflect set-based dissimilarity, with Dice-Sørensen varying more smoothly. 
Overlap coefficient emphasizes shared elements relative to the smaller set, yielding lower distances when overlaps exist.


\subsubsection{Distance Analysis} 
\label{distance_analysis}
We identify differences by analyzing the distribution of likely invariant distances at each breakpoint. 
To systematically detect changes, we adopt Kernel Density Estimation (KDE) for its robustness and model-agnostic simplicity. 
KDE smoothes and estimates the probability density of distances from multiple inputs, enabling distribution visualization without assuming an underlying model~\cite{Silverman1986DensityAnalysis}. 
 It has been effectively used in prior studies~\cite{Bosu2021AnalyzingDatasets, dErrico2021AutomaticClustering} to capture and analyze variations in data distributions.
 In particular, we detect local maxima, i.e., points in the distance distribution higher than their immediate neighbors, indicating distinct peaks in the data for each measure.
A breakpoint is deemed behaviorally shifted, i.e., \textit{flagged}, when at least two of four distance measures show non-zero local maxima in their KDE distributions.
We opt to use a 50\% threshold to ensure consistent signals emerge across multiple measures.
Although we do not claim the best performance, this approach should regulate false alarms and improve statistical agreement and reliability even when individual measures are imperfect~\cite{Pease1980ReachingFaults, Lamport1998TheParliament}.
Indeed, the threshold can be adjusted to fit the target needs.
The KDE computations are performed using the \texttt{KernelDensity} module from the \texttt{scikit-learn} Python package~\cite{Buitinck2013APIProject}, version 1.3.2.

%% file: section/4-study.tex
This section outlines our research questions and explains the experimental setup and fuzzing configuration.

\subsection{Research Questions}
In this work, we aim to understand the potential of~\ourFramework{} in identifying behavioral changes that may indicate bugs.
We answer the following research questions.

\inlineheading{RQ1) To what extent do vulnerability-related bugs produce different likely invariants compared to the clean version?  }
Vulnerability, which can cause severe consequences~\cite{Stecklein2004ErrorCycle, Karg2011AResearch}, may evade fuzzing without suitable oracles.
We aim to investigate how often likely invariant differences arise in vulnerabilities.
To answer this question, we compare likely invariants in clean and buggy versions of vulnerabilities in Magma~\cite{Hazimeh2020Magma:Benchmark}.

\inlineheading{RQ2) How effectively can \ourFramework{} detect real-world regression bugs? }
Building on RQ1, we want to further understand the effectiveness of likely invariant analysis to detect behavioral changes in regression bugs, a type of bug caused by mistakes during code changes that occur frequently in software development~\cite{Zhu2021RegressionFuzzing} and potentially lead to vulnerabilities.
To answer this question, we apply \ourFramework{} to flag buggy program points in regression bugs in BugOSS~\cite{Kim2024BugOss:Techniques}.

\inlineheading{RQ3) Can \ourFramework{} outperform SAST warnings for bug detection in regression code changes? 
}
Prior work~\cite{Charoenwet2024AnReview} shows that SAST warnings can help detect functions containing vulnerable code changes. 
As \ourFramework{} shares a similar goal, we want to compare it against SAST to evaluate whether likely invariant analysis can offer complementary support for code review. 
To answer this question, we use SAST detection rates as a baseline on the same dataset used in RQ2.



\subsection{Datasets}
\label{experiment_dataset_preparation}
To facilitate our experiments, the dataset must include fuzz drivers, i.e., entry points that enable fuzzing tools to generate inputs for the target programs.
We identify two datasets that meet our requirements.
For RQ1, we use Magma~\cite{Hazimeh2020Magma:Benchmark} -- a well-known semi-synthetic fuzzing benchmark dataset, comprising 120 front-ported bugs of various 
difficulties.
For RQ2, we use BugOSS~\cite{Kim2024BugOss:Techniques} -- a manually curated real-world dataset, comprising 21 code commits containing real-world regression bugs from OSS-Fuzz~\cite{OSS-FuzzOSS-Fuzz}, the continuous fuzzing platform for open-source projects. 

We select Magma for RQ1 because it offers a large set of vulnerabilities with systematic support for creating \textit{buggy} and \textit{clean} program versions.
For RQ2 and RQ3, we select BugOSS, which contains code commits that introduced regression bugs (\textit{buggy} version), providing real-world \textit{buggy} and \textit{benign} changes, along with the prior commits that represent the \textit{clean} version.
These commits closely resemble those encountered during quality assurance in CI/CD pipelines, making them appropriate for evaluating \ourFramework{}.

\input{table/table-magma-dataset}

Although we want to use all target bugs in the selected datasets, we need to refine the selection to ensure a consistent and comparable evaluation setting across all cases.
For Magma, we exclude \texttt{SQLite} because the amalgamation process merges all source files, obstructing breakpoint setup.
For BugOSS, we exclude \texttt{gdal-47716} and \texttt{poppler-35789} because they employ a multi-stage build process, affecting the framework integration.
We also exclude \texttt{yara-38952} because it introduces new functionality in the buggy version, resulting in the breakpoint being unavailable in the clean version.
Information on the selected bugs in Magma (118) and BugOSS (18) is shown in Table~\ref{table:magma-dataset} and Table~\ref{table:bugoss-dataset}.
\input{table/table-bugoss-dataset}

\subsection{Experimental Setup}
\label{experimental_setup}
For RQ1, we set breakpoints for the modified code block in the Magma bug patches.
Specifically, we automatically set a \textit{subsequent} breakpoint after every modified code block.
At each breakpoint, we extract the in-scope local variables and function arguments, i.e., variables accessible at the program's execution point within the function, using the \texttt{frame variable} command in LLDB~\cite{LLDBMap}.
To manage complexity, we extract up to ten variables per breakpoint. 
Too many variables can lead to an overwhelming number of invariants, diluting meaningful differences, hindering analysis, and increasing computational cost.
Note that we do not set a breakpoint after the Canary\footnote{Magma's monitoring instrumentation} because it is not a part of the bug and may inject biases into likely invariant analysis.
In total, 183 breakpoints are set for 118 bugs in eight target programs in RQ1.


For RQ2, we select breakpoints that reflect the impact of code changes in the buggy version. 
Given the bug-inducing lines are known, 
breakpoints are chosen using a set of rules applied in order.
\begin{enumerate}
    \item select the last line in the code block if the code change is a statement, allowing value propagation
    \item select the last line of the inner block if the code changes cover a code block 
    \item select the condition line if the code changes include a code block that only has a return statement, to ensure observation within the function's control flow
\end{enumerate}

At each breakpoint, we select variables that are present in both versions and affected by the actual code changes, specifically those that are referred to or assigned within the modified code.
To enable cross-version comparison, we exclude breakpoints that are present in only one version of the program or that reference variables not shared between versions.
To evaluate possible false alarms, we include breakpoints after code changes unrelated to the target bugs (i.e., \textit{benign} changes), applying the same criteria for variable selection.
In total, we set 47 breakpoints for 18 bugs in 18 target programs in RQ2.

As regression bugs can be relevant to refactorings~\cite{Bavota2012WhenStudy}, we annotate whether selected bug-introducing changes are likely refactorings, using definitions from~\cite{Murphy-Hill2012HowIt}.
In general, we label a bug as likely refactoring if the change preserves behavior and involves only structural modifications, such as code reorganization without explicit functional changes.

\begin{figure}[t]
  \centering
  \includegraphics[width=0.9\linewidth, scale=1]{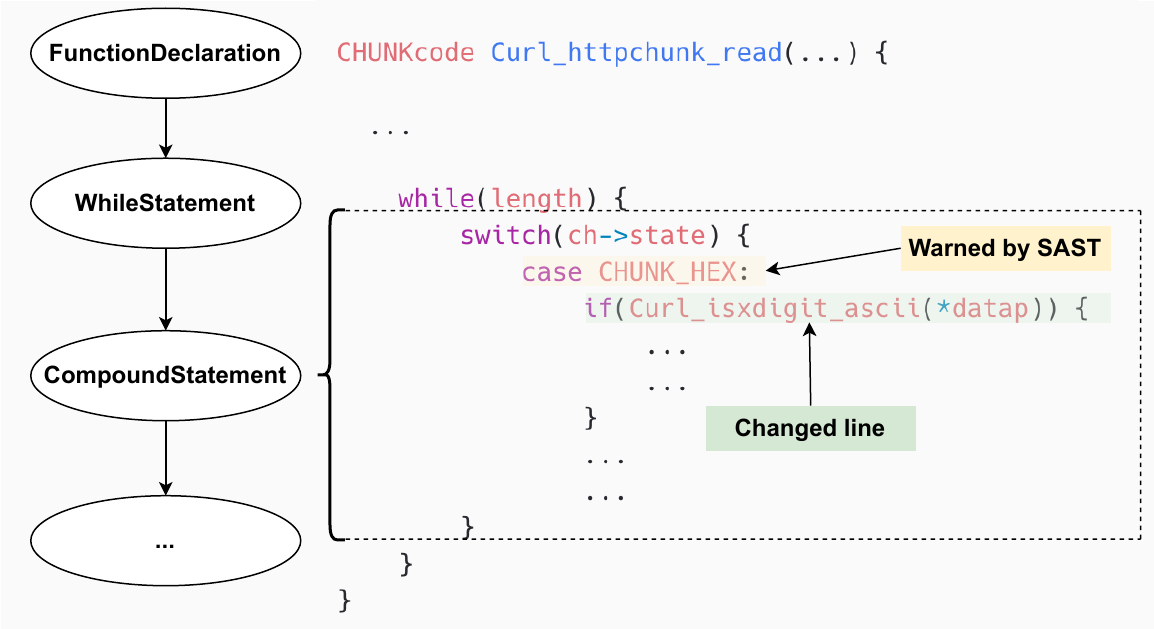} 
  \caption{Mapping SAST warnings with changed code block using Abstract Syntax Tree (BugOSS: \textit{curl-8000})}
  \label{fig:sast-ast-mapping}
\end{figure}

For RQ3, following~\cite{Charoenwet2024AnReview}, we run three top-performing SAST tools: Flawfinder, CodeChecker, and CodeQL on the buggy versions in BugOSS and match their warnings to code changes.
Although prior work~\cite{Charoenwet2024AnReview} shows SAST warnings can support function-level bug detection, achieving finer-grained level detection remains challenging.
As warnings may appear on different lines~\cite{Lipp2022AnDetection}, we focus on block-level rather than line-level detection.
To support block-level evaluation, we generate abstract syntax trees (ASTs) of the changed functions using \texttt{tree-sitter}~\cite{Tree-sitter} and map each warning to the compound statement enclosing the code change.
For example, a warning manifested in a loop can be associated with the entire loop construct as shown in Fig.~\ref{fig:sast-ast-mapping}.
This mapping facilitates the evaluation of SAST detection and likely invariant differences at the same block level as the selected breakpoints.

\subsection{Fuzzing Setup}
\label{fuzzing_experiment}
To fuzz the target programs and generate inputs, we follow the settings used by the datasets~\cite{Hazimeh2020Magma:Benchmark, Kim2024BugOss:Techniques}.
We select three fuzzing tools to represent diverse techniques and capabilities, offering a broad perspective on methodologies commonly used in real-world scenarios.
AFL++~\cite{TheAFLplusplus} is a state-of-the-art fuzzing tool that identified numerous real-world critical vulnerabilities.
Honggfuzz~\cite{HonggfuzzHonggfuzz} is an open-source fuzzing tool developed by Google that outperforms other tools on the Magma benchmark.
libFuzzer~\cite{LibFuzzerDocumentation} is the lightweight built-in fuzzing tool of the Clang compiler.
We use up-to-date fuzzing tools, specifically AFL++ v4.21c, Honggfuzz v2.6, and libFuzzer in Clang-15.


We use the build scripts, fuzz drivers, and initial seeds from each dataset.
We build two program versions per target: clean and buggy. 
If there are multiple fuzz drivers, we select the one with the highest coverage. 
For Magma, coverage data for \texttt{openssl} and \texttt{php} is based on OSS-Fuzz reports, while for \texttt{libxml2}, we ran a coverage test due to missing data. 
BugOSS provides a single fuzz driver per program.

The buggy version of each target program is fuzzed for 36 core-hours (six CPU cores for six hours) on a virtual machine running Ubuntu 20.04.
For BugOSS, we apply the Change-aware Seed Reuse (CSR)~\cite{Yoo2021ImprovingHANA} technique to increase the chance that fuzzing inputs reach the changed functions, based on BugOSS's recommendation~\cite{Kim2024BugOss:Techniques}.
If CSR fails to generate inputs that reach the changed functions in the clean version, we fuzz the buggy version with initial seeds. 
We then collect inputs that do not trigger crashes, i.e., \textit{non-crashing inputs}, from the input queue for the next steps.
If no inputs reach the changed functions, we manually run the program to verify binary correctness and repeat fuzzing to rule out methodological misconfigurations

%% file: table/table-magma-dataset.tex
\begin{table}[h]
\centering
\scriptsize
\caption{Magma Dataset for RQ1--bug information in selected targets}

\begin{tabularx}{\linewidth}{
    >{\hsize=1\hsize}X |
    >{\centering\arraybackslash\hsize=1\hsize}X
    >{\centering\arraybackslash\hsize=1\hsize}X 
    >{\centering\arraybackslash\hsize=1\hsize}X
}
\toprule
\textbf{Subject} & \textbf{Total Bugs }&\textbf{ Hard-to-Fuzz Bugs\textsuperscript{\textdagger} }& \textbf{Breakpoints } \\ 
 \midrule
\textbf{libpng} & 7 & 2 & 13 \\ 
\textbf{libsndfile} & 18 & 0 & 20 \\ 
\textbf{libtiff} & 14 & 3 & 27 \\ 
\textbf{libxml2} & 17 & 3 & 28 \\ 
\textbf{lua} & 4 & 0 & 12 \\ 
\textbf{openssl} & 20 & 5 & 26 \\ 
\textbf{php} & 16 & 2 & 29 \\ 
\textbf{poppler} & 22 & 5 & 28 \\ 
 \bottomrule
\multicolumn{4}{l}{\textdagger~Bugs that do not manifest after 24-hour fuzzing campaign reported by Magma~\cite{Hazimeh2020Magma:Benchmark}}
 
\end{tabularx}

\label{table:magma-dataset}
\vspace*{-\baselineskip}
\end{table}

%% file: table/table-bugoss-dataset.tex
\begin{table}[h]
\centering
\scriptsize
\caption{BugOSS Dataset for RQ2 \& RQ3--selected bugs with number of changes in three granularities and refactoring-related label}

\begin{tabularx}{\linewidth}{
    >{\hsize=1.8\hsize}X|
    >{\centering\arraybackslash\hsize=0.7\hsize}X 
    >{\centering\arraybackslash\hsize=0.8\hsize}X|
    >{\centering\arraybackslash\hsize=0.7\hsize}X
    >{\centering\arraybackslash\hsize=1\hsize}X
}
\toprule
\textbf{Subject} & \textbf{Changed Files} & \textbf{Changed Functions} & \textbf{Breakpoints} & \textbf{Refactoring-Related\textsuperscript{\textdagger}} \\
 \midrule
\textbf{arrow-40653} & 1 & 2 & 3 & Yes \\
\textbf{aspell-18462} & 1 & 1 & 1 & Yes \\
\textbf{curl-8000} & 3 & 2 & 2 & Yes \\
\textbf{exiv2-50315} & 3 & 3 & 1 & Yes \\
\textbf{file-30222} & 1 & 3 & 1 & No \\
\textbf{grok-28418} & 3 & 10 & 4 & No \\
\textbf{harfbuzz-55779} & 3 & 4 & 2 & No \\
\textbf{leptonica-25212} & 1 & 1 & 2 & No \\
\textbf{libarchive-44843} & 1 & 5 & 3 & No \\
\textbf{libhtp-17198} & 1 & 2 & 9 & No \\
\textbf{libxml2-17737} & 1 & 1 & 1 & Yes \\
\textbf{ndpi-49057} & 1 & 2 & 3 & Yes \\
\textbf{openh264-26220} & 1 & 1 & 1 & No \\
\textbf{openssl-17715} & 3 & 6 & 3 & No \\
\textbf{pcapplusplus-23592} & 3 & 6 & 2 & No \\
\textbf{readstat-13262} & 1 & 1 & 2 & No \\
\textbf{usrsctp-18080} & 1 & 1 & 1 & Yes \\
\textbf{zstd-21970} & 5 & 8 & 6 & No \\
 \bottomrule
\multicolumn{5}{l}{\textdagger~Changes involve structural modifications and likely preserving behaviors}
 
\end{tabularx}

\label{table:bugoss-dataset}
\end{table}

%% file: section/5-result.tex
We report the experiment results and answer the research questions in this section.

\subsection{RQ1: The differences of likely invariant in vulnerabilities}
\label{magma_detection_experiment}
We use \ourFramework{} to analyze likely invariants at breakpoints following the modified code in Magma (Section~\ref{experimental_setup}).
We compare likely invariants at each breakpoint across the \textit{buggy} and \textit{clean} versions, i.e., the target program with and without injected bugs.
Likely invariants are generated using non-crashing inputs from AFL++, Honggfuzz, and libFuzzer (Section~\ref{fuzzing_experiment}).
We consider a vulnerability bug to have likely invariant differences if \ourFramework{} flags at least one breakpoint in such a bug.
We report the results and findings below.


\inlineheading{Results:} 
Table~\ref{table:rq3-magma-result} shows the number of bugs for which \ourFramework{} flags likely invariant differences.
In total, 10\%-58\% of Magma bugs show likely invariant differences in one or more breakpoints.
The non-crashing input corpus from three fuzzing tools reached 25\%-75\% of the bugs. 
Many flagged bugs with likely invariant differences share common traits, such as having slight changes in a statement or conditions.
For example, \texttt{TIF010} (\texttt{AAH018} -- CVE-2018-8905), flagged by \ourFramework{}, involves a modification of a while loop condition.

We observe that the absence of likely invariant differences in some subjects may be attributed to the nature of Magma patches.
After an investigation, we find that code changes preceding over 45\% of the breakpoints introduce fixes, e.g., condition checks.
This suggests that buggy versions often lack necessary safeguards, rather than containing explicit buggy code.
As a result, the two versions may exhibit similar behavior around these breakpoints, limiting observable differences.
For example, \texttt{SSL016} (\texttt{MAE111} -- CVE-2015-1788) has a conditional break in the clean version, which is difficult to trigger via fuzzing, that is missing in the buggy version.

\input{table/table-rq3-magma_result}


\textbox{\textbf{Finding 1.1 Vulnerability's likely invariants:}  
10\%-58\% of the vulnerability bugs exhibit likely invariant differences.
}


Several bugs flagged by \ourFramework{} correspond to hard-to-fuzz bugs that \textit{survived}, or remained untriggered, in the Magma benchmark~\cite{Hazimeh2020Magma:Benchmark}, despite at least 24 hours of fuzzing, a duration commonly adopted in many fuzzing studies and projects.
We find that 
up to 80\% of the hard-to-fuzz bugs in one subject (\texttt{poppler}) are flagged in at least one breakpoint.
Hard-to-fuzz bugs often require multiple conditions to be satisfied, making it challenging for fuzzers.
For example, \texttt{PNG004} (\texttt{AAH004} -- CVE-2015-0973) exhibits likely invariant differences due to a variable being cast between data types, though the bug is only triggered under some exact input values.
Likely invariant analysis can capture the effects of this behavior.

\textbox{\textbf{Finding 1.2 Hard-to-fuzz bugs’ likely invariants:} Up to 80\% of bugs not triggered after a 24-hour fuzzing campaign exhibit differences in likely invariants.}





\subsection{RQ2: \ourFramework{} detecting regression bugs }
\label{bugoss_detection_experiment}
To answer RQ2, 
we apply \ourFramework{} to BugOSS, which represents real-world regression code changes.
The \textit{buggy} version is the regression commit, while the \textit{clean} version is the code commit prior to the regression commit.
We select breakpoints and variables (Section~\ref{experimental_setup}), fuzz the buggy version of the program (Section~\ref{fuzzing_experiment}), and analyze the likely invariant difference.
We report the number of bugs that \ourFramework{} flags in at least one buggy breakpoint.

\inlineheading{Results:} 
\ourFramework{} successfully analyzes likely invariants in 16 out of 18 bugs in the BugOSS dataset.
It flags at least one buggy breakpoint in 75\% of the reached bugs, as shown in Table~\ref{table:rq1-bugoss-result}. 
Notably, over 41.2\% (5 out of 12) of the flagged bugs 
are associated with refactoring changes (Table~\ref{table:bugoss-dataset}) where most variables remain unchanged, but the surrounding logic has been modified.
For example, \texttt{aspell-18462} replaces a while loop with a for loop using an incorrect exit condition, which is detectable through likely invariants.

\input{table/table-rq1-bugoss_result}

\textbox{\textbf{Finding 2.1 Detecting regression bugs:} \ourFramework{} flags likely invariant differences in 75\% of fuzzing-reached bugs. }

\begin{figure}[t]
  \centering
  \includegraphics[width=0.7\linewidth, scale=1]{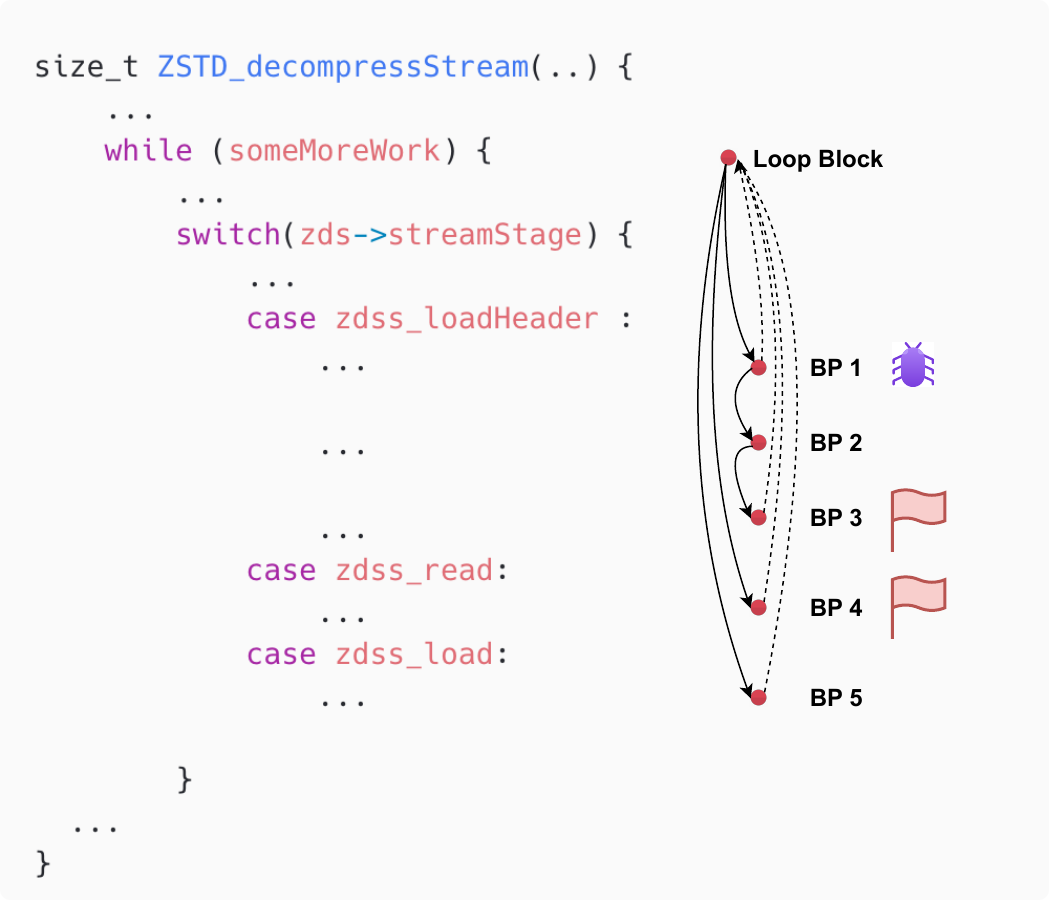} 
  \caption{Interdependency between breakpoints via the program's control-flow can cause likely invariant differences in non-buggy breakpoints (BP3-BP4)}
  \label{fig:rq2-false-alarm}
\end{figure}

\needAttention{
Table~\ref{table:rq1-bugoss-result} shows that \ourFramework{} flags at least one benign breakpoint in 41.2\% of the flagged bugs in our dataset (5 out of 12).
These flagged benign breakpoints constitute false alarms that may distract reviewers from identifying the truly buggy breakpoints.
Our investigation reveals that most false alarms arise from the program’s data-flow and control-flow dependencies.
For example, non-buggy breakpoints involve the variables modified by buggy code changes, allowing propagation of faulty data.
Fig.~\ref{fig:rq2-false-alarm}. illustrates the control-flow between breakpoints in \texttt{zstd-21970}, showing that the variables in a buggy breakpoint can propagate to other breakpoints.
By accounting for such interdependent breakpoints through simple dependency tracing, i.e., treating all breakpoints on the same call graph as a single unit, false alarms can be decreased to 16.7\% of the bugs (2 out of 12).
}
These remaining false alarms may stem from intentional behavioral changes, although this remains inconclusive due to limited information about the specific changes in each commit.

\textbox{\textbf{Finding 2.2 False alarms and handling strategies:} 
41.7\% of bugs are flagged in non-buggy breakpoints, partly due to the data propagation. 
Accounting for interdependencies can help reduce false alarms, in our study down to 16.7\%.
}

We examine the false negatives of \ourFramework{} by reviewing the cases where no buggy breakpoints were flagged.
Some inputs reached the selected breakpoints and successfully generated likely invariants, but still could not flag four bugs, i.e., \texttt{arrow-40653}, \texttt{exiv2-50315}, \texttt{file-30222}, and \texttt{readstat-13262}.
Based on our results, we identify two primary reasons why some bugs are not flagged. 
First, the only conditions that can trigger likely invariant differences may also be detected by fuzzing tools, cases that fall outside the scope of \ourFramework{}. 
For example, \texttt{file-30222} bug was due to a missing null check. 
Non-crashing inputs may lack the necessary variability to reveal these issues. 
Second, when the buggy code is confined to a single statement, it may not offer sufficient behavioral complexity to motivate exploration of all edge cases by fuzzing tools once overall paths are met~\cite{Boehme2021Fuzzing:Reflections, Liang2018Fuzzing:Art}.


\textbox{\textbf{Finding 2.3 Limitations:} \ourFramework{} is limited when bugs can only surface by very specific, or crashing, inputs or where fuzzing fails to thoroughly explore input space.}

\subsection{RQ3: Evaluating \ourFramework{} for code review}
\label{bugoss_hit_rate_experiment}
We compare \ourFramework{} results with SAST in flagging possible bugs in code changes.
A study~\cite{Charoenwet2024AnReview} shows that identifying buggy changed functions with SAST warnings can significantly reduce code review effort.
To this end, we compare RQ2 results against the SAST warnings via the hit rates on the buggy code at various granularities.
Similar to prior works~\cite{Lipp2022AnDetection, Charoenwet2024AnReview}, we determine the detection performance based on various scenarios.
A bug is considered \textit{detected} by SAST warnings at the file, function, or code block level when at least one warning is produced in buggy files, functions, or code blocks.
Specifically, at the code block level, we associate a warning with a block using AST-based mapping, as explained in Section~\ref{experimental_setup}.
On the other hand, a bug is considered \textit{detected} by \ourFramework{} at the file, function, or code block level when at least one breakpoint is flagged in any buggy file, function, or code block.
We also report false alarms when SAST produces warnings on, or \ourFramework{} flags, only non-buggy (benign) code in a given bug.
We consider \ourFramework{}'s hit rate of each fuzzing tool to highlight their individual performance.

\begin{figure}[t]
  \includegraphics[width=\linewidth, scale=1]{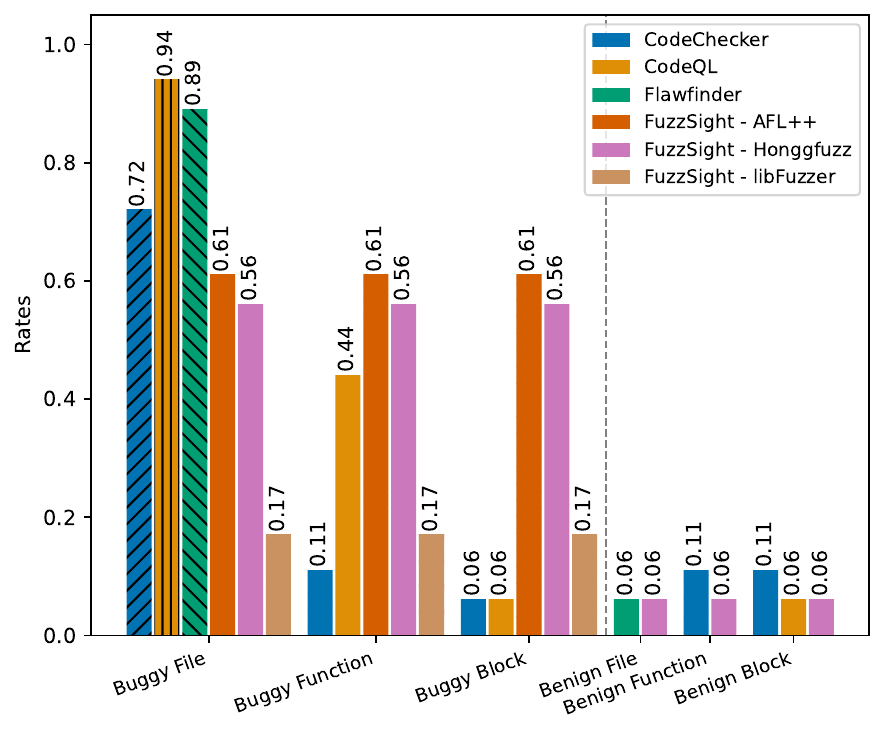} 
  \caption{Hit rates and False alarms--number of bugs for which SAST warnings and \ourFramework{} correctly flagged at least one buggy code, or mistakenly flagged all benign code, at different granularities  }
  \label{fig:rq2-hit-rate}
\end{figure}

\inlineheading{Results:} 
\ourFramework{} can detect more buggy code blocks compared to SAST warnings.
Fig.~\ref{fig:rq2-hit-rate}. displays the hit rates in the three granularity-based scenarios. 
SAST warnings are capable of detecting buggy changed files in 72\%-94\% of the bugs.
At the function level, the SAST warning detection rates drop to 11\%-44\%.
At the block level, SAST warnings can detect only 6\% of the bugs.
On the other hand, \ourFramework{} achieves 17\%–61\% detection rate throughout all scenarios.

\textbox{\textbf{Finding 3.1 Block-level hit rate:} \ourFramework{} flags buggy code blocks in 61\% of the bugs, while SAST warnings appear in buggy code blocks of only 6\% of the bugs.}

False alarms can mislead bug detection among code changes. 
While SAST warnings are effective at detecting buggy files and functions, false alarms also occur in 6\%–11\% of bugs where warnings only appear in non-buggy files, functions, and code blocks.
\ourFramework{} correctly flags the buggy breakpoint in nearly all bugs being flagged, with only one exception case where Honggfuzz mistakenly flagged a non-buggy breakpoint, affecting up to 6\% of the total bugs.

\textbox{\textbf{Finding 3.2 Comparing False Alarms with SAST:} SAST produces warnings in non-buggy changed blocks for 6\%–11\% of the bugs, while \ourFramework{} does so in only 6\%.}


%% file: table/table-rq3-magma_result.tex
\begin{table}[t]
\centering
\scriptsize
\caption{RQ1 Results--number of bugs for which \ourFramework{} flags likely invariant differences in at least one breakpoint using non-crashing inputs from three fuzzing tools}

\begin{tabularx}{\linewidth}{
    >{\hsize=0.8\hsize}X |
    >{\centering\arraybackslash\hsize=1\hsize}X |
    >{\centering\arraybackslash\hsize=1\hsize}X
    >{\centering\arraybackslash\hsize=1.2\hsize}X
}
\toprule
\textbf{Subject} & 
\textbf{Reached Bugs} & 
\textbf{Flagged Bugs}  & 
\textbf{Flagged Hard-to-fuzz Bugs} \\
 \midrule
\textbf{libpng} & 5 (71\%) & 3 (43\%) & 1 (50\%)   \\
\textbf{libsndfile} & 8 (44\%) & 2 (11\%)   & -   \\
\textbf{libtiff}  & 7 (50\%)  & 7 (50\%)   & 1 (33\%)   \\
\textbf{libxml2}  & 10 (41\%)  & 10 (58\%)   & 2 (67\%)   \\
\textbf{lua}  & 3 (75\%) & 2 (50\%)   & -  \\
\textbf{openssl} &  5 (25\%) & 2 (10\%)   & 0 (0\%)   \\
\textbf{php} & 4 (25\%)  & 3 (19\%)   & 0 (0\%)   \\
\textbf{poppler} & 14 (64\%) & 8 (36\%)   & 4 (80\%)  \\
\bottomrule
\end{tabularx}

\label{table:rq3-magma-result}
\end{table}

%% file: table/table-rq1-bugoss_result.tex
\begin{table}[t]
\centering
\scriptsize
\caption{RQ2 Results--numbers of reached and flagged breakpoints are accumulated from inputs of three fuzzing tools}

\begin{tabularx}{\linewidth}{
    >{\hsize=2.2\hsize}X|
    >{\centering\arraybackslash\hsize=.6\hsize}X
    >{\centering\arraybackslash\hsize=.6\hsize}X|
    >{\centering\arraybackslash\hsize=.7\hsize}X
    >{\centering\arraybackslash\hsize=.7\hsize}X
    >{\centering\arraybackslash\hsize=1.2\hsize}X
}
\toprule
\textbf{Subject} & \textbf{Total BPs} & \textbf{Buggy BPs} & \textbf{Reached BPs} & \textbf{Flagged BPs} & \textbf{Buggy BPs Flagged} \\
 \midrule
 \multicolumn{6}{l}{\textbf{Flagged Bugs}} \\
 \midrule
\textbf{aspell-18462} & 1 & 1 & 1 & 1 & 1 \\
\textbf{curl-8000} & 2 & 1 & 1 & 1 & 1 \\
\textbf{grok-28418} & 9 & 2 & 8 & 4 & 1 \\
\textbf{harfbuzz-55779} & 5 & 2 & 5 & 1 & 1 \\
\textbf{libhtp-17198} & 2 & 1 & 2 & 2 & 1 \\
\textbf{libxml2-17737} & 3 & 1 & 3 & 2 & 1 \\
\textbf{ndpi-49057} & 5 & 3 & 4 & 2 & 2 \\
\textbf{openh264-26220} & 1 & 1 & 1 & 1 & 1 \\
\textbf{openssl-17715} & 2 & 2 & 2 & 2 & 2 \\
\textbf{pcapplusplus-23592} & 6 & 1 & 6 & 4 & 1 \\
\textbf{usrsctp-18080} & 1 & 1 & 1 & 1 & 1 \\
\textbf{zstd-21970} & 5 & 1 & 3 & 3 & 1 \\
\midrule
\multicolumn{6}{l}{\textbf{Missed Bugs}} \\
\midrule
\textbf{arrow-40653} & 2 & 2 & 1 & 0 & 0 \\
\textbf{exiv2-50315} & 2 & 1 & 2 & 0 & 0 \\
\textbf{file-30222} & 1 & 1 & 1 & 0 & 0 \\
\textbf{readstat-13262} & 1 & 1 & 1 & 0 & 0 \\
\midrule
\multicolumn{6}{l}{\textbf{Unreached Bugs}} \\
\midrule
\textbf{leptonica-25212} & 3 & 1 & 0 & 0 & 0 \\
\textbf{libarchive-44843} & 5 & 3 & 0 & 0 & 0 \\
 \bottomrule
\end{tabularx}

\label{table:rq1-bugoss-result}
\end{table}

%% file: section/6-discussion.tex
{\color{Black}

We discuss the implications of our findings for the potential use of FuzzSight in code review in this section. 
We frame the discussion around four dimensions: intended use, benefits for code review, operational considerations, and future work.

\subsection{Intended Use of \ourFramework{}}
A central motivation of \ourFramework{} is to empower reviewers to act as an active \textit{intelligent oracle}, interpreting behavioral signals.
Existing review support often leans on syntactic changes or SAST warnings, which risk overlooking semantic discrepancies. 
By surfacing likely invariant differences between program versions, \ourFramework{} could highlight the potential behavioral changes that may otherwise remain hidden, providing an additional lens for the reviewers.

Our experiments demonstrate that \ourFramework{} detects meaningful invariant shifts associated with a substantial portion of vulnerability-inducing bugs (10\%–58\%) and in up to 80\% of bugs untriggered after long fuzzing campaigns (RQ1). 
These findings indicate that behavioral differences are relatively common in buggy code and that non-crashing fuzzing inputs provide a practical way to expose them.
In this way, \ourFramework{} can serve as a complementary signal in code review: alongside syntactic checks or automated warnings, it offers reviewers additional behavioral context to help judge whether a detected difference reflects an intentional change or a potential defect.


Nonetheless, these signals are not absolute indicators of bugs.
A detected difference may reflect a deliberate semantic refinement (as illustrated in our preliminary experiments) or a defect (such as a regression in data structure handling).
Interestingly, we observed that most refactoring-related bugs (5 out of 7 in our dataset, RQ2) were among the flagged cases, suggesting that unintended structural changes frequently manifest as behavioral shifts in likely invariants.
The value of \ourFramework{} lies in providing such behavioral context, not in replacing the reviewer’s judgment.
In this way, \ourFramework{} serves as a supporting information provider, bridging the gap between abstract specifications and concrete execution traces.

\subsection{Complementary Benefits of \ourFramework{} in Code Review}
We now discuss the possible application of \ourFramework{} in practice.
In particular, we outline several ways in which the findings suggest potential benefits.

\inlineheading{Complementarity with existing tools.}
In regression bugs, \ourFramework{} reports invariant differences in 75\% of fuzzing-reached bugs (RQ2). 
At the block level, it highlights buggy blocks in 61\% of the bugs, while SAST warnings cover only 6\% (RQ3). 
This disparity illustrates how static and dynamic analyses capture different aspects of code quality. 
Likely invariants reflect runtime semantics, whereas SAST relies on syntactic or structural properties. 
Combined, they provide a more comprehensive context for code review.

\inlineheading{Precision relative to SAST.}
\ourFramework{} demonstrates higher precision at the block level compared to SAST. 
While SAST warnings appear in buggy code blocks for only 6\% of the bugs and in non-buggy changed blocks for 6\%–11\%, \ourFramework{} flags buggy blocks in 61\% of the bugs and produces false positives in non-buggy changed blocks for only 6\% (RQ3).

Considering all flagged breakpoints, \ourFramework{}’s false positives are higher (41.7\% of bugs, RQ2) due to dynamic propagation, where an upstream change can affect downstream breakpoints. 
However, accounting for interdependencies through lightweight graph analyses reduces these distractions to 16.7\% of bugs, highlighting that many false alarms are systematic and can be mitigated. 
This distinction underscores that, at the granularity relevant for review (changed blocks), \ourFramework{} offers both higher hit rates and fewer false alarms compared to SAST, while the remaining propagated signals can be managed with simple dependency understanding that are a part of code review practices~\cite{Goncalves2025CodeTheories}.


\subsection{Operational Considerations}
\label{operationl_consideration}
Adopting \ourFramework{} in practice requires attention to several operational factors that influence its effectiveness and usability. 
In this subsection, we discuss key considerations for configuring and deploying the framework.

\inlineheading{Fuzzing budget and campaign time.} 
\ourFramework{} relies on execution traces, which in turn depend on fuzzing campaigns. 
While 24-hour campaigns are standard in fuzzing research, review cycles often demand shorter feedback loops~\cite{Kudrjavets2022MiningAnalysis}.
Our experiments suggest that even shorter runs can yield informative signals (RQ1  \& RQ2), but longer runs can also uncover additional invariant differences, as illustrated by \texttt{file-30222}. 
This highlights a tension as deeper fuzzing yields richer behavioral signals, but may exceed practical review timelines. 
A potential compromise is to integrate \ourFramework{} into CI pipelines, where longer campaigns accumulate traces incrementally, while reviewers access the latest available signals during review.


\inlineheading{Configuration, Scalability, and Interpretability.}
The effectiveness of \ourFramework{} depends on careful configuration of breakpoints, variable inclusion, and timeout thresholds, as these choices can influence both precision and computational cost. 
Overly broad instrumentation may flood reviewers with irrelevant signals, while overly narrow selections risk missing critical differences. 
Our experiments demonstrate that lightweight dependency analyses can help filter false alarms (RQ2), reducing distractions while retaining coverage.

Even with the computational overhead of likely invariant mining, \ourFramework{} generates useful signals at a scale suitable for typical code changes (RQ2 \& RQ3).
By carefully managing the number and scope of traced breakpoints and variables, the framework ensures that signals remain understandable and relevant, allowing reviewers to focus on meaningful behavioral changes while minimizing false positives. 
This emphasizes that scalability, configuration, and signal interpretability are closely tied to workflow context, ensuring that the framework’s outputs are both usable and informative for code review purposes.

One promising direction is adaptive configuration, where breakpoints are prioritized based on contextual information such as historical bug patterns or other review signals, further improving efficiency without compromising coverage.



\inlineheading{Tool Independence and Input Generation Flexibility.}
\ourFramework{} requires external tools to generate dynamic execution traces.
In our implementation, fuzzing is used to produce diverse program inputs, enabling likely invariant mining and detection of behavioral shifts. 
While fuzzing provides practical coverage for our experiments, the framework is not inherently tied to any specific input generator. 
Any sufficiently rich input pool or alternative generator that exercises relevant program paths can be substituted, allowing \ourFramework{} to adapt to different languages, domains, or testing workflows. 
In fact, as illustrated in RQ3 (Fig.~\ref{fig:rq2-hit-rate}.), different fuzzing tools exhibit varying effectiveness, underscoring that the framework’s utility stems from the availability of diverse inputs rather than reliance on a particular tool.
This modularity ensures that the framework’s effectiveness is not limited by the choice of fuzzing tool and can evolve alongside advances in input generation techniques.

}

\subsection{Future Work}
We discuss potential directions to extend \ourFramework{}.

\inlineheading{Address limitation in execution trace generation.}
Execution traces cannot capture program elements beyond the variable's values.
For example, a bug in an inline statement with no variable assignment, such as in \texttt{poppler-35789}~\footnote{\url{https://gitlab.freedesktop.org/poppler/poppler/-/commit/2b2808719d2c91283ae358381391bb0b37d9061d}}, can go undetected.
As likely invariants depend on execution traces, enhancing the trace generation process is crucial.

This challenge is compounded by \ourFramework{}'s reliance on fuzzing, requiring fuzzing infrastructure such as fuzz drivers for input generation.
If fuzzing saturates coverage, important likely invariants may not be inferred. 
\needAttention{
Moreover, it is ineffective for bugs that manifest only under highly specific inputs or under certain crash conditions. }
Addressing these limitations will require improvements in both \ourFramework{} and fuzzing tools.

\inlineheading{Investigate automated solutions for false alarm reduction.}
In RQ2, false alarms from interdependent breakpoints are reduced by identifying their relationships, which can be done via taint analysis, call graphs, or control-flow graphs.
However, different techniques are suitable for different projects. 
Similarly, the sensitivity for detecting differences in KDE distributions (Section~\ref{distance_analysis}), a potential source of false negatives, can also be adjusted.
We leave this exploration for future work.

\inlineheading{Investigate code change prioritization.}
Static application security testing (SAST) tools can help prioritize code changes~\cite{Charoenwet2024AnReview}.
In contrast, fuzzing tools usually detect concrete failures, emphasizing soundness over completeness~\cite{Barr2015TheSurvey}, limiting their utility for prioritization.
In RQ3, we show that likely invariant analysis using fuzzing inputs can help flag buggy code changes, potentially supporting prioritization.
However, our dataset includes limited code changes, making it difficult to draw definitive conclusions. 
Further investigation is necessary.

\inlineheading{Address performance overhead.}
Post-fuzzing trace generation is a repetitive process. 
Integrating trace generation into a fuzzer can streamline data generation and reduce overhead.
On the other hand, mining likely invariants from execution traces remains time-intensive.
We observe that processing a single execution trace can take several minutes, becoming impractical when analyzing large inputs. 
To ensure the feasibility of this study, we applied a timeout for the mining process.
Despite practicality, it highlights the need for scalable techniques that balance the output quality and the applicability.

%% file: section/8-threats.tex
\inlineheading{Internal Validity}
Building the executable binary is a complex process. 
Project-specific setup may be incorrect due to missing information or some changes in the build scripts.
To reduce this threat, we rely on the scripts provided by the datasets and the recent version of OSS-Fuzz to initiate the experiments. 

We fuzzed target programs for 36 core-hours, following prior work~\cite{Kim2024BugOss:Techniques}.
However, fuzzing durations may affect outcomes.
We replicated selected experiments using a 24-hour campaign and discuss the results in Section~\ref{operationl_consideration}.
Finally, 
as some breakpoints can record over 20k execution traces, we limit each breakpoint to 1,500 randomly sampled traces to
prevent memory issues. 
While this may impact analysis quality, the sample size supports a 99\% confidence interval with a margin of error under 0.05, ensuring reliable results.

\inlineheading{Construct Validity}
Construct validity concerns whether the chosen measurements accurately capture the intended concept.
We acknowledge that the distance measures used in this study are not perfect.
We selected four distance measures previously applied in software engineering research~\cite{Allyson2019SherlockCode, Shtern2012ClusteringEngineering, Taheri2020Similarity-basedFeatures}, to capture different types of differences in likely invariant sets.
Other distance measures may be suitable for certain subjects.

\inlineheading{External Validity}
External validity concerns the generalizability of the results.
We are aware that our approach may not be able to cover every type of bug.
While our datasets include 136 vulnerabilities and regression bugs from over 20 diverse OSS projects,
our results may not necessarily reflect all types of vulnerabilities and bugs.

%% file: section/9-conclusion.tex
This work investigates the potential of dynamic program analysis techniques to enhance code reviews.
We develop \ourFramework{} to analyze likely invariant differences between two program versions and detect unexpected behaviors using non-crashing fuzzing inputs.
We test \ourFramework{} with Magma and BugOSS datasets, which represent synthetic vulnerabilities and real-world bug-introducing code commits.
The results show that \ourFramework{} flags 10\%-58\% vulnerabilities and 75\% of regression bugs reached by fuzzing inputs.
Notably, up to 80\% of hard-to-fuzz vulnerabilities can be flagged.
We further evaluate the detection rates of \ourFramework{} against a baseline of SAST warnings.
\ourFramework{} correctly flags ten times more bugs than SAST warnings at the code block level, with lower false alarm rates.
These findings suggest that \ourFramework{} can detect behavioral changes at the code-block level, complementing SAST warnings, which are more cost-effective at the file and function levels.
They highlight the potential of combining fuzzing and invariant analysis to bring dynamic insights into early code review.

%% file: section/10-acknowledgements.tex
This work was supported by the use of Nectar Research Cloud, a collaborative research platform supported by NCRIS-funded Australian Research Data Commons (ARDC). 
Patanamon Thongtanunam and Van-Thuan Pham were supported by two Australian Research Council’s Discovery Early Career Researcher Award (DECRA) projects (DE210101091 and DE230100473).